\documentclass[12pt]{article}
\usepackage{amssymb}
\usepackage{amsmath}
\usepackage{epsf}
\usepackage[mathscr]{eucal}
\title{Primordial fluctuations in bulk inflaton model}
\author{Kazuya Koyama, Keitaro Takahashi \\
Department of Physics, University of Tokyo \\
7-3-1 Hongo, Bunkyo, Tokyo 113-0033, Japan}

\begin{document}
\maketitle

\begin{abstract}
An inflationary brane model driven by a bulk inflaton with exponential 
potential is proposed. We find a family of exact solutions that describe
power-law inflation on the brane. These solutions enable us to derive  
exact solutions for metric perturbations analytically. By calculating 
scalar and tensor perturbations, we obtain a spectrum of primordial 
fluctuations at 
the end of the inflation. The amplitudes of scalar and tensor 
perturbations are enhanced in the same way if
the energy scale of the inflation is sufficiently higher than the 
tension of the brane. Then the relative amplitude of scalar and 
tensor perturbations is not suppressed even for high-energy inflation. 
This is a distinguishable feature from the inflation model driven by 
inflaton on the brane where tensor perturbations are suppressed for 
high-energy inflation.  We also point out that massive Kaluza-Klein 
modes are not negligible at high-frequencies on 3-space of our 
brane. 

\end{abstract}

\section{Introduction}

Higher dimensional formulation is necessary for theories such as 
superstring and M theory. All these theories are confronted
with a problem that our universe seems to be four-dimensional, which
is conventionally coped with compactifications of the extra dimensions.
Braneworld model \cite{HoravaWitten,RSI,RSII} where our universe is a four-dimensional 
subspace in a higher dimensional spacetime, is an epoch-making scenario because
it broke the conventional idea that the extra dimensions must be compact and small. 
In this scenario, ordinary matter fields are confined on the brane, while graviton 
can propagate through the extra dimensions. Thus higher-dimensional feature of 
spacetime must emerge in the nature of gravity.

In the early universe, inflation can amplify vacuum fluctuation of Kalzua Klein modes 
as well as a massless mode. Thus inflation is a powerful tool to probe the extra dimensions.
Inflation can occur by the dynamics of inflaton either on the 
brane or in the bulk. The former model has been investigated by many authors
\cite{Kaloper,Nihei,Kim,KoyamaSoda,GarrigaSasaki,Hawking,NO,MWBH,Langlois,MizunoMaeda}
and it has been known that inflation occurs basically in the same way as the
ordinary four-dimensional universe. On the other hand, there have been rather few
study of the latter model. One of the difficulties is that one should treat the dynamics
of the bulk inflaton including the back-reaction to the bulk geometry.
Kobayashi et al. \cite{KobayashiKoyamaSoda} dealed with
a bulk inflaton by using slow-roll approximation, which corresponds to increasing
bulk cosmological constant effectively. 
Himemoto et al. \cite{Himemoto1,Himemoto2,Sago,Himemoto3,Himemoto4} included the dynamics of
an inflaton in the bulk perturbatively in terms of $Hl <1 $ and $\vert m \vert l<1$ 
where $H$ is the Hubble scale on the brane, $m$ is the inflaton mass and $l$ is the 
curvature radius of the Anti-de Sitter bulk, respectively. They showed that 
the inflaton in the bulk can mimic the 4D inflaton dynamics at low energies.
However, the dynamics of the inflaton at high energies is still unsolved. 
Another difficulty is the analysis of the
perturbations in this model. In order to calculate the 
curvature perturbation generated in the inflation, we should consistently 
take into account metric perturbations. In the 5D spacetime with 
scalar field, it is quite a complicated issue. Hence only the perturbation
of the inflaton was considered by neglecting metric perturbations. 
So far, to our knowledge, there has been no progress on this issue. 

In this paper we propose a new model where dynamics of inflaton can be 
exactly solved including the back-reaction to the bulk spacetime even at high-energies. 
We deal with a bulk scalar field with exponential potential.
This type of scalar field arises from a sphere reduction in M theory or string
theory\cite{Cvetic}. Cosmological solutions in this model were investigated in
Ref \cite{Lukas,SetoKodama,OchiaiSato}. In Ref.\cite{KobayashiKoyama2002},
it was pointed out that various inflationary scenarios can be realized in this 
model. In this paper, we consider a inflation model driven by bulk 
potential energy. A remarkable thing is that the metric perturbations in this 
model can be exactly solved. Then we can consistently 
calculate the primordial fluctuations 
and examine a feature of bulk inflaton model 
which should be compared with conventional 4D models or inflation models driven by
inflaton on the brane. 
Because the calculation of the perturbations
involves rather lengthy derivations, we do not present detailed analysis
here. Complete treatment of cosmological perturbations is presented in an 
accompanied paper \cite{KoyamaTakahashi}.  

This paper is organized as follows. In section \ref{section:background} we
give a family of background solution. Based on this, we calculate cosmological 
perturbations and show that both scalar and tensor perturbations are described
by 5D massless scalar field in section \ref{section:perturbation}.
In section \ref{section:massless}, we investigate the quantum fluctuations of 5D massless 
field. A spectrum of primordial fluctuations at the end of the inflation 
is presented in section \ref{section:spectrum}. Section \ref{section:discussion}
is devoted to discussions.

\section{Background spacetime\label{section:background}}

We start from the five-dimensional Einstein-Hilbert action with
a bulk scalar field,
\begin{equation}
S = \int d^{5}x \sqrt{- g_{5}} \left( \frac{1}{2 \kappa^{2}} R 
- \frac{1}{2} \partial_{\mu} \phi \partial^{\mu} \phi - \Lambda(\phi) \right)
- \int d^{4}x \sqrt{- g_{4}} \lambda(\phi),
\end{equation}
where $\kappa^{2}$ is five-dimensional gravitational constant. 
The potential of the scalar field in the bulk and on the brane are taken to be exponential:
\begin{eqnarray}
\kappa^{2} \Lambda(\phi) & = & \left( \frac{\Delta}{8} + \delta \right) \lambda^{2}_{0}
e^{-2 \sqrt{2} b \kappa \phi}, \\ 
\kappa^{2} \lambda(\phi) & = & \sqrt{2} \lambda_{0} e^{- \sqrt{2} b \kappa \phi}.
\end{eqnarray}
Here $\lambda_{0}$ is the energy scale of the potential, $b$ is the dilaton coupling 
and we defined 
\begin{equation}
\Delta = 4 b^{2} - \frac{8}{3}.
\end{equation}
We assume the $Z_2$ symmetry across the brane.

For $\delta=0$, the static brane solution was found \cite{Cvetic}. 
The existence of the static brane requires tunning between bulk potential
and brane tension known as Randall-Sunrum tunning.
It has been shown that for $\Delta \leq -2$, we can avoid the presence of the naked 
singularity in the bulk and also ensure the trapping of the gravity. The reality of the dilaton
coupling requires $-8/3 \leq \Delta$. For $\Delta=8/3$, we recover Randall-Sundrum
soltuion. 
The value of $\delta$, which is not necessarily small, represents a deviation from the 
Randall-Sundrum tuning. This deviation yields an inflation on the brane.

We assume a solution of the form;
\begin{eqnarray}
ds^{2} &=& e^{2 W(z)} \left( - dt^{2} + e^{2 \alpha(t)} \delta_{ij} dx^{i} dx^{j}
+ e^{2 \sqrt{2}b \kappa \phi(t)} dz^{2} \right),\nonumber\\
\phi(t,z) &=& \phi(t) + \Xi(z).
\end{eqnarray}
Then five-dimensional Einstein equations and the equation of motion for the scalar field 
give the equation for $W(z)$ and $\Xi(z)$;
\begin{equation}
 3(W'' + {W'}^{2}) + \frac{1}{2} \kappa^{2} {\Xi'}^{2} +
\left(\frac{\Delta}{8} + \delta \right) \lambda_{0}^{2} e^{2W - 2 \sqrt{2} b \kappa \Xi}
+ \sqrt{2} \lambda_{0} e^{W - \sqrt{2} b \kappa \Xi} \delta(z - z_{0}) 
= -\lambda_0^2 \frac{\Delta+4}{\Delta} \delta,
\label{eq:Einstein_ttz}
\end{equation}
\begin{equation}
6 {W'}^{2} - \frac{1}{2} \kappa^{2} {\Xi'}^{2} 
+ \left(\frac{\Delta}{8} + \delta \right) \lambda_{0}^{2} e^{2W - 2 \sqrt{2} b \kappa \Xi}  
= \lambda_0^2 \delta,
\label{eq:Einstein_zzz}
\end{equation}
\begin{equation}
\Xi'' + 3 W' \Xi' + 2 \sqrt{2} \frac{b}{\kappa} 
\left( \frac{\Delta}{8} + \delta \right) \lambda_{0}^{2} e^{2W - 2 \sqrt{2} b \kappa \Xi} 
+ 2 \frac{b}{\kappa} \lambda_{0}  e^{W - \sqrt{2} b \kappa \Xi}  \delta(z - z_{0}) 
= -4 \sqrt{2} \frac{b}{\kappa} \frac{\delta}{\Delta} \lambda_0^2,
\label{eq:EOMz}
\end{equation}
where prime denotes the derivative with respect to $z$.
Thanks to the Bianchi identity, one of the above three equations
are not independent. 
Equation for $\alpha(t)$ and $\phi(t)$ are also obtained as 
\begin{equation}
\dot{\alpha}^2+\sqrt{2}b \kappa \dot{\phi} \dot{\alpha}=
\frac{1}{6} \kappa^2 \dot{\phi}^2-\frac{1}{3}\lambda_0^2 \frac{\Delta+4}{\Delta}
\delta e^{-2 \sqrt{2}b \kappa \phi},
\end{equation}
\begin{equation}
\ddot{\alpha}+2 \dot{\alpha}^2 +\frac{1}{6} \kappa^2 \dot{\phi}^2
=\frac{1}{3} \lambda_0^2 \delta e^{-2 \sqrt{2} b \kappa \phi},
\end{equation}
\begin{equation}
\ddot{\phi}+(3 \dot{\alpha} + \sqrt{2} b \kappa \dot{\phi}) \dot{\phi}
=-4 \sqrt{2} b \kappa^{-1} \lambda_0^2 \frac{\delta}{\Delta} e^{-2 \sqrt{2}b \kappa \phi},
\end{equation}
where dot denotes the derivative with respect to $t$. 
Again one of three equations are not independent. 

The solution for $\alpha(t)$ and $\phi(t)$ can be easily found as 
\begin{eqnarray}
e^{\alpha(t)} &=& (H_0 t)^{\frac{2}{3 \Delta+8}}=(- H \eta)^{\frac{2}{3(\Delta+2)}} ,
\label{eq:solution_alpha}
\\
e^{ \sqrt{2} b \kappa \phi(t)} &=& H_0 t=\left( - H \eta \right)^{\frac{3 \Delta + 8}{3 (\Delta + 2)}}.
\label{eq:solution_phi}
\end{eqnarray}
where 
\begin{equation}
H_0 \equiv -\frac{3\Delta+8}{3 (\Delta +2)} H, \quad 
H=- (\Delta + 2) \sqrt{- \frac{\delta}{\Delta}} \lambda_{0},
\end{equation}
and a conformal time $\eta$ is defined as 
\begin{equation}
\eta = \int e^{-\alpha} dt
= \frac{3 \Delta + 8}{3 (\Delta + 2)} H_{0}^{-\frac{2}{3 \Delta + 8}} 
t^{\frac{3 (\Delta + 2)}{3 \Delta + 8}}.
\end{equation}
We should notice that power-law inflation occurs on the brane for $-8/3 < \Delta < -2$.
Thus in the rest of the paper we shall assume $-8/3 < \Delta < -2$.

The solutions for $W(z)$ and $\Xi(z)$ can be written as 
\begin{equation}
e^{W(z)} = {\cal H}(z)^{\frac{2}{3(\Delta+2)}},\quad e^{\kappa \Xi(z)}=
{\cal H}(z)^{\frac{2 \sqrt{2}b}{(\Delta+2)}},
\end{equation}
where the form of ${\cal{H}}(z)$ depends on the sign of $\Delta/8 + \delta$.
According to the sign of
$\Delta/8 + \delta$, solutions for ${\cal H}(z)$ are obtained as follows:

(1)$\frac{\Delta}{8} + \delta < 0$
\begin{equation}
{\cal H}(z)= \sqrt{-1-\frac{\Delta}{8 \delta}} \sinh{H z},
\label{eq:background_negative} 
\end{equation}

(2)$\frac{\Delta}{8} + \delta = 0$
\begin{equation}
{\cal H}(z)=e^{H z}
\label{eq:background_zero} 
\end{equation}

(3)$\frac{\Delta}{8} + \delta > 0$
\begin{equation}
{\cal H}(z)= \sqrt{1+\frac{\Delta}{8 \delta}} \cosh{H z},
\label{eq:background_positive} 
\end{equation}
At the location of the brane $z=z_0$
the solutions should satisfy junction conditions;
\begin{equation}
\partial_z W(z) \vert_{z=z_0}=-e^{W(z_0)-\sqrt{2}b \kappa \Xi(z_0)} 
\frac{\sqrt{2}}{6} \lambda_0,\quad 
\partial_z \Xi(z) \vert_{z=z_0}=-  e^{W(z_0)-\sqrt{2}b \kappa \Xi(z_0)}
b \kappa^{-1} \lambda_0.
\end{equation}
Then the location of the brane is determined by ${\cal H}(z=z_0)=1$.
In the following sections, we assume
$\Delta/8+\delta \leq 0$ for which the potential energy in the bulk 
is negative or zero.

\section{Cosmological perturbations \label{section:perturbation}}

In this section we consider perturbations in this background.
The perturbations can be decomposed into scalar, vector and tensor 
perturbations in terms of their properties on the 3-space
at fixed $t$ and $y$ coordinate. 
Taking appropriate gauge fixing conditions, the perturbed metric
and scalar field is give by
\begin{eqnarray}
ds^2 &=& e^{2 W(z)} \left[ e^{2 \sqrt{2}b \kappa \phi(t)}(1+2 N)dz^2 
+2 A  dt dz -(1+2 \Phi )dt^2+e^{2 \alpha(t)}(1-2 \Psi )
\delta_{ij} dx^i dx^j \right], \nonumber\\
\phi &=& \phi(t)+\Xi(z)+\delta \phi,
\end{eqnarray}
for scalar perturbations and
\begin{equation}
ds^2=e^{2 W(z)} \left( e^{2 \sqrt{2} b \kappa \phi(t)} dz^2 
-dt^2 + e^{2 \alpha(t)} \left(-2 T_i dz dx^i
-2 S_i dt dx^i+ \delta_{ij} dx^i dx^j \right) \right),
\end{equation}
for vector perturbations and 
\begin{equation}
ds^{2} = e^{2 W(z)} \left[
e^{2 \sqrt{2} b \kappa \phi}dz^2 - dt^{2} + e^{2 \alpha(t)}
\left( \delta_{ij} + h_{ij}\right) dx^{i} dx^{j}\right],
\end{equation}
for tensor perturbations.
Here $S_i$ and $T_i$ are transverse vector 
($\nabla^i S_i=0$ and $\nabla^i T_i=0$) and 
$h_{ij}(t,x,z)$ is transverse and traceless tensor($h^i_i=0,
\nabla^i h_{ij}=0$) 
where $\nabla_{i}$ is the derivative operator on 3-space $\delta_{ij}$.
Although 5D Einstein equations for these perturbations are
rather complicated, the solutions for these perturbations
can be analytically obtained. Because the derivation is very long, 
the detailed analysis will be presented in an accompanied paper 
\cite{KoyamaTakahashi}
and we only present important results here.

Introducing canonical variable for scalar perturbations 
\begin{equation}
\delta \phi_f \equiv \delta \phi+\frac{\dot{\phi}}{\dot{\alpha}} \Psi
=\delta \phi+3 \sqrt{2} b \kappa^{-1} \Psi,
\end{equation}
the second-order perturbed 5D action for gravity plus scalar field
for scalar perturbations can be written as 
\begin{equation}
\delta S^{(S)}=\frac{1}{2}\int dz dt dx^3 e^{3 W(z)} e^{\sqrt{2} b \kappa \phi(t)}
e^{3 \alpha(t)}(e^{-2 \sqrt{2}b \kappa \phi(t)} \delta \phi_f^{'2}- \dot{\delta \phi_f}^2
+e^{-2 \alpha(t)} (\nabla \delta \phi_f)^2).
\end{equation}
The junction condition for canonical variable $\delta \phi_f$ is 
given by
\begin{equation}
\partial_z \delta \phi_f \vert_{z=z_0} =0.
\end{equation}
In a similar way, the canonical variables for tensor perturbation is given by
\begin{equation}
\varphi_{ij}=\frac{1}{2 \kappa} h_{ij}.
\label{eq:tensor}
\end{equation}
Then the second-order perturbed 5D action for tensor perturbations is obtained as
\begin{equation}
\delta S^{(T)}=\frac{1}{2}\int dz dt dx^3 e^{3 W(z)} e^{ \sqrt{2} b \kappa \phi(t)}
e^{3 \alpha(t)}(e^{-2 \sqrt{2} b \kappa \phi(t)} \varphi_{ij}^{'2}- \dot{\varphi_{ij}}^2
+e^{-2 \alpha(t)} (\nabla \varphi_{ij})^2).
\end{equation}
The junction condition is given by 
\begin{equation}
\partial_z \varphi_{ij} \vert_{z=z_0} =0.
\end{equation}
These actions are nothing but the action for 5D massless scalar field
where tensor perturbation has two degrees of freedom due to polarization.
Hence both scalar perturbation and tensor perturbation are described by 
5D massless scalar field with Neunmann boundary condition at the barne. 
As in a conventional 4D inflationary cosmology, these perturbations are quantum 
mechanically generated during inflation. Then in the next section, we will 
investigate the quantum fluctuations of the 5D massless scalar field in this 
background spacetime. 

Finally we comment on vector perturbations. In conventional 4D Einstein 
gravity, vector perturbation $S_i$ is constrained to vanish for vanishing matter
vorticity. In the brane world, the bulk vector perturbation $T_i$ can support
the vector perturbations on the brane even for vanishing matter vorticity
\cite{Vector}.  
However, in this background, we can show that the vector perturbations have no 
normalizable zero-mode and do not affect the perturbations at large scales. 
Thus we do not consider the vector perturbations in this paper. 

\section{Quantization of massless scalar field \label{section:massless}}
In the previous section, the metric perturbations were shown to be described
by a 5D massless scalar field. Thus, we shall investigate the quantum 
fluctuation of the massless scalar field in this background spacetime.
The calculations are essentially the same as those in Ref. \cite{KobayashiKoyamaSoda}
where they considered the case for $\Delta=-8/3$ (see also Ref. \cite{Sago}).
 
The equation of motion for massless field  $\varphi(t,x,z)$ is
\begin{equation}
e^{2 \sqrt{2} b \kappa \phi} \left[ \ddot{\varphi} 
+ (3 \dot{\alpha} + \sqrt{2} b \kappa \dot{\phi}) \dot{\varphi}
- \nabla_{k}\nabla^{k} \varphi\right]
= \varphi'' +  3 W' \varphi'.
\end{equation}
We can use the separation of variables to solve this equation and expand $\varphi(t,x,z)$ as,
\begin{equation}
\varphi(t,x,z) = \int dm d^{3}p \left[a_{pm} \psi_{m}(z) \chi_{m}(t) e^{i p x}
+ ({\rm h.c.}) \right]
\end{equation}
Here $a_{pm}$ is the annihilation operator and satisfies the following
commutation relation,
\begin{equation}
\left[ a_{pm}, a_{p'm'}^{\dagger} \right] = \delta(p - p') \delta(m - m').
\end{equation}
On the other hand, $\psi_{m}(z)$ and $\chi_{m}(t)$ satisfy the following equations,
\begin{eqnarray}
&& \psi''_{m} + 3 W' \psi' + m^{2} \psi = 0, \label{eq:psi} \\
&& \ddot{\chi}_{m} + (3 \dot{\alpha} + \sqrt{2} b \kappa \dot{\phi}) \dot{\chi}_{m}
+ (e^{-2 \alpha} p^{2} + e^{-2 \sqrt{2} b \kappa \phi} m^{2}) \chi = 0,
\label{eq:chi}
\end{eqnarray}
where $m$ corresponds to the Kaluza-Klein mass.
 In the following subsections, we discuss the solutions for
Eq. (\ref{eq:psi}) and Eq. (\ref{eq:chi}) and their properties.

\subsection{Mode function $-$ z-direction $-$}

Because of $Z_{2}$ symmetry along the z-direction, the junction condition for
$\psi_{m}(z)$ is
\begin{equation}
\left. \frac{\partial}{\partial z} \psi_{m}(z) \right|_{z = z_{0}} = 0.
\end{equation}
With this condition and the commutation relation for $\varphi$, the solution for
Eq. (\ref{eq:psi}) is completely determined.
In order to know the behavior of the solutions, it is convenient to define
\begin{equation}
\psi_{m}(z) = e^{-3 W(z)/2}f_m(z) .
\end{equation}
Then, Eq. (\ref{eq:psi}) gives the Schr\"{o}dinger-like equation for
$f_m(z)$
\begin{equation}
- f_m'' + V_{eff}(z) f_m = m^{2} f_m.
\end{equation}
The effective potential is given by
\begin{eqnarray}
V_{eff}(z) & = & \frac{3}{2} W'' + \frac{9}{4} {W'}^{2} \nonumber \\
& = & - \frac{\Delta + 1}{\Delta} \left( \frac{\Delta}{8} + \delta \right) \lambda_{0}^{2}
{\cal H}(z)^{-2} - \frac{1}{\Delta} \delta \lambda_{0}^{2}
- \frac{1}{\sqrt{2}} \lambda_{0} \delta(z - z_{0}),
\label{eq:effective}
\end{eqnarray}
where we used the solutions for $W(z)$. The term proportional to 
${\cal H}(z)^{-2}$ arises from the negative potential energy in the bulk.
The explicit form of the potential is determined by the background spacetime but
it can be seen that there is zero mode confined on the brane irrespective of the
background spacetime. We consider the following two cases:

\begin{figure}[t]
\epsfxsize=11.3cm
\centerline{\epsfbox{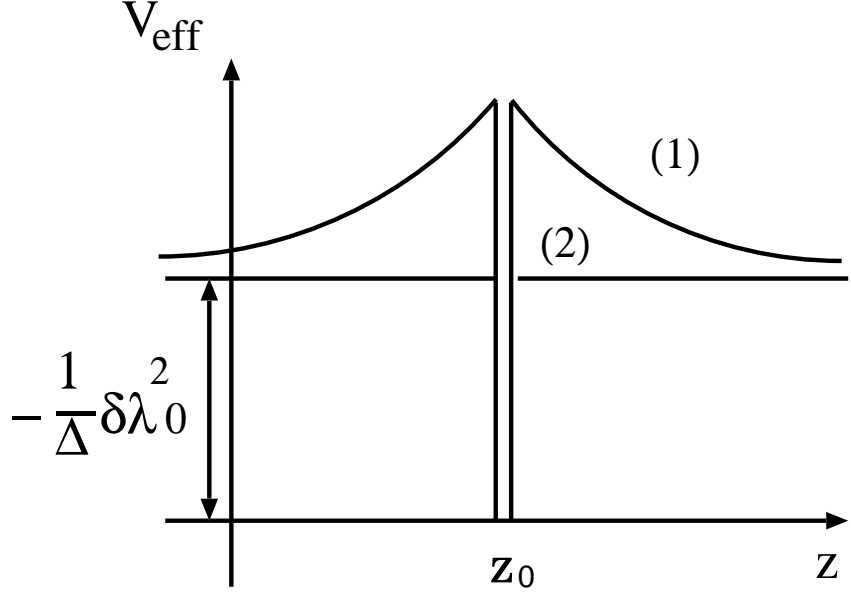}}
\caption{Effective potential $V_{eff}$. (1) and (2) correspond to
the case that $\Delta/8 + \delta$ is negative and zero respectively.}
\label{fig:volcano}
\end{figure}

\hspace{1cm}\\
(1)$\frac{\Delta}{8} + \delta < 0$\\
From (\ref{eq:background_negative}), the effective potential $V_{eff}$ is
obtained as
\begin{equation}
V_{eff}(z) = \frac{\Delta + 1}{\Delta} (\sinh{H z})^{-2} \lambda_{0}^{2}
- \frac{1}{\Delta} \delta \lambda_{0}^{2}
- \frac{1}{\sqrt{2}} \lambda_{0} \delta(z - z_{0}),
\end{equation}
and is shown in Fig. \ref{fig:volcano}. 

The equation for $\psi_m(z)$ becomes
\begin{equation}
\psi''_{m} + \frac{2 H}{\Delta + 2} \coth{H z} \psi'_{m} + m^{2} \psi_{m} = 0.
\end{equation}
There are zero mode and continuum massive modes with 
$m^{2} \geq H^{2}/(\Delta + 2)^{2} = - \delta \lambda_{0}^2 / \Delta$.
Massive modes with $0 < m^{2} < H^{2}/(\Delta + 2)^{2}$ are not normalizable and
are not investigated here. The solutions for zero mode and normalizable massive modes are
\begin{eqnarray}
\psi_{0}(z) & = & \frac{1}{\sqrt{2}} 
\left( - 1 - \frac{\Delta}{8 \delta} \right)^{- \frac{1}{2(\Delta + 2)}}
\left[ \int^{\infty}_{z_{0}} (\sinh{Hz})^{\frac{2}{\Delta + 2}} dz \right]^{- \frac{1}{2}},
\label{eq:C0} \\
\psi_{m}(z) & = & \sqrt{\frac{H}{2}} 
\left( - 1 - \frac{\Delta}{8 \delta} \right)^{- \frac{1}{2(\Delta + 2)}}
(|\xi|^{2} + |\zeta|^{2})^{- \frac{1}{2}} (\sinh{H z})^{- \mu} \nonumber \\
&& \times \left[ P^{\mu}_{\nu}(\cosh{Hz}) 
+ C Q^{\mu}_{\nu}(\cosh{H z}) \right],
\end{eqnarray}
where $P^{\mu}_{\nu}$ and $Q^{\mu}_{\nu}$ are associated Legendre functions and
\begin{eqnarray}
\mu & = & - \frac{\Delta}{2(\Delta + 2)}, \quad
\nu  =  - \frac{1}{2} + i \beta, \quad
\label{eq:mu}
\beta  =  \sqrt{\frac{m^{2}}{H^{2}} - \frac{1}{(\Delta + 2)^{2}}}, \\
\xi & = & \frac{\Gamma(i \beta)}{\Gamma(\frac{\Delta + 1}{\Delta + 2} + i \beta)}, \quad
\zeta  =  \frac{\Gamma(-i \beta)}{\Gamma(\frac{\Delta + 1}{\Delta + 2} - i \beta)}
+ C \pi e^{\mu \pi i} \frac{\Gamma(\frac{1}{\Delta + 2} + i \beta)}{\Gamma(1 + i \beta)}, 
\label{eq:xi}\\
C & = & - \frac{(\frac{\Delta + 1}{\Delta + 2} + i \beta) P^{\mu}_{\nu + 1}(\cosh{H z_{0}})
- (\frac{1}{\Delta + 2} + i \beta) \cosh{H z_{0}} P^{\mu}_{\nu}(\cosh{H z_{0}})}
{(\frac{\Delta + 1}{\Delta + 2} + i \beta) Q^{\mu}_{\nu + 1}(\cosh{H z_{0}})
- (\frac{1}{\Delta + 2} + i \beta) \cosh{H z_{0}} Q^{\mu}_{\nu}(\cosh{H z_{0}})}.
\end{eqnarray}

\hspace{1cm}\\
(2)$\frac{\Delta}{8} + \delta = 0$\\
From (\ref{eq:background_zero}), the effective potential $V_{eff}$ is
obtained as
\begin{equation}
V_{eff}(z) = - \frac{1}{\Delta} \delta \lambda_{0}^{2}
- \frac{1}{\sqrt{2}} \lambda_{0} \delta(z - z_{0}),
\label{eq:nobarrier}
\end{equation}
and is shown in Fig. \ref{fig:volcano}.
The equation for $\psi_m(z)$ becomes
\begin{equation}
\psi''_{m} +\frac{2}{\Delta+2} H \psi'_{m} + m^{2} \psi_{m} = 0.
\end{equation}
Again, massive modes with $0 < m^{2} < H^{2}/(\Delta + 2)^{2}$ are not normalizable. 
The solutions for zero mode and normalizable massive modes are
\begin{eqnarray}
\psi_{0}(z) & = & \sqrt{-\frac{H}{\Delta+2}}, \\
\psi_{m}(z) & = & \sqrt{\frac{H}{4 \pi}} 
\left( \exp \left[\left(-\frac{1}{\Delta+2} - i \beta \right) H z \right] 
+ C \exp \left[\left(-\frac{1}{\Delta+2} + i \beta \right) H z \right]
\right),
\end{eqnarray}
where $\beta$ is the same as Eq. (\ref{eq:mu}) and  
\begin{equation}
C=\frac{i \beta + \frac{1}{\Delta+2}}{i \beta - \frac{1}{\Delta+2}}.
\end{equation}
In this case, we can easily calculate the ratio of $\vert \psi_m \vert^2$ 
to $\vert \psi_0 \vert^2$ on the brane;
\begin{equation}
\frac{\vert \psi_m(z_0) \vert^2}{\vert \psi_0(z_0) \vert^2}=
\frac{-(\Delta+2)}{\pi} \left(\frac{\frac{m^2}{H^2}-\left(\frac{1}{\Delta+2} \right)^2}
{\frac{m^2}{H^2}} \right).
\label{eq:ratio}
\end{equation}

\subsection{Mode function $-$ time-direction $-$}

From Eq. (\ref{eq:solution_alpha}) and Eq. (\ref{eq:solution_phi}), 
Eq. (\ref{eq:chi}) becomes
\begin{equation}
\ddot{\chi}_{m} + \frac{3 \Delta+14}{3 \Delta+8} t^{-1} \dot{\chi}_{m}
+ \left[ p^{2} \left( H_{0} t \right)^{- \frac{4}{3 \Delta+8}} 
+ \frac{m^{2}}{H_{0}^{2}} t^{-2} \right] \chi_{m} = 0.
\end{equation}
We choose the Bunch-Davis vacuum, which coincides with the Minkowski vacuum at high 
frequencies. With this condition and the commutation relation, the solution to
Eq. (\ref{eq:chi}) is completely determined.
The solutions for zero mode and normalizable massive modes with 
$m^{2} \geq H^{2}/(\Delta + 2)^{2}$ are
\begin{eqnarray}
\chi_{0}(\eta) & = & \frac{\sqrt{\pi}}{2} H^{-\frac{1}{2}} (- H \eta)^{- \frac{1}{\Delta + 2}}
H_{- \frac{1}{\Delta + 2}}^{(1)}(- p \eta), \\
\chi_{m}(\eta) & = & \frac{\sqrt{\pi}}{2} H^{-\frac{1}{2}} (- H \eta)^{- \frac{1}{\Delta + 2}}
e^{- \frac{\beta \pi}{2}} H_{i \beta}^{(1)}(- p \eta),
\end{eqnarray}
where
\begin{equation}
\beta = \sqrt{\frac{m^{2}}{H^{2}} - \frac{1}{(\Delta + 2)^{2}}}.
\end{equation}

\subsection{Vacuum expectation value of fluctuations}

Both zero mode and massive mode contribute to the spectrum and 
the vacuum expectation value of massless scalar field will be,
\begin{equation}
\langle \varphi^{2}(x) \rangle = 
\int d^{3}p \left[ |\psi_{0}(z_0)|^{2} |\chi_{0}|^{2}
+ \int^{\infty}_{-\frac{H}{\Delta + 2}} dm \; |\psi_{m}(z_0)|^{2} |\chi_{m}|^{2} \right].
\end{equation}
But this naive formula is not correct because in the large $m$ and $p$ limit 
$|\chi_{m}|^{2}$ behaves
\begin{equation}
|\chi_{m}|^{2} \sim \frac{(- H \eta)^{3}}{2 H \sqrt{(m/H)^{2} + (- p \eta)^{2}}}
\;\;\;\; (m,p \rightarrow \infty),
\end{equation}
and the $m$-integral diverges logarithmically. This ultra-violet divergence appears
in the five-dimensional field theory even in Minkowski spacetime. Thus we have to
subtract this divergence as following:
\begin{equation}
|\chi_{m}|^{2} - \frac{(- H \eta)^{3}}{2 H \sqrt{(m/H)^{2} + (- p \eta)^{2}}}.
\end{equation}
Then the correct expectation value is,
\begin{eqnarray}
\langle \varphi^{2}(x) \rangle & = & 
\int d^{3}p \left[ |\psi_{0}(z_0)|^{2} 
\left( |\chi_{0}|^{2} - \frac{(- H \eta)^{3}}{2 H \sqrt{(- p \eta)^{2}}} \right) \right. 
\nonumber \\
&& + \left. \int^{\infty}_{-\frac{H}{\Delta + 2}} dm \; |\psi_{m}(z_0)|^{2} 
\left( |\chi_{m}|^{2} - \frac{(- H \eta)^{3}}{2 H \sqrt{(m/H)^{2} + (- p \eta)^{2}}} \right)
\right] \\
& \equiv & \int d^{3}p \frac{2 \pi^2}{p^{3}} 
\left[ \langle \varphi^{2} \rangle_{0} + \langle \varphi^{2} \rangle_{m} \right],
\end{eqnarray}
where $\langle \varphi^{2} \rangle_{0}$ and $\langle \varphi^{2} \rangle_{m}$ are
contribution from zero mode and massive mode, respectively.
In the small $\eta$ limit, each contribution can be written as,
\begin{eqnarray}
\langle \varphi^2 \rangle_{0} 
& \longrightarrow & C_{0}^{2} \lambda_0 \left( \frac{H}{2 \pi} \right)^{2} 
\left( \frac{p}{H} \right)^{\frac{3 \Delta + 8}{\Delta + 2}}, 
\label{eq:massless} \\
\langle \varphi^2 \rangle_{m}  
& \longrightarrow & C_{m}^{2} \lambda_0 \left( \frac{H}{2 \pi} \right)^{2} 
 \left( \frac{p e^{\alpha(\eta)}}{H} \right)^{3}.
\end{eqnarray}

In order to interpret the results, it should be noted that 
bulk curvature scale and Hubble constant on the brane are 
determined by the bulk potential and the deviation from the RS tuning, respectively.
Thus their ratio,
\begin{equation}
r = \left| \frac{\delta}{\Delta/8 + \delta} \right|,
\label{eq:HL}
\end{equation}
determines the behavior of $\langle \varphi^2 \rangle$. For small $r$, we expect
4D physics is recovered because the horizon scale is larger
than the scale of the extra-dimension. On the other hand for large $r$, 
the effects of the bulk could be large. In addition, $r$ determines 
the effective size of the extra dimension defined by
\begin{equation}
L(t) \equiv \int^{\infty}_{z_0} dz e^{3W(z)} e^{\sqrt{2} b \kappa \phi_0(t)}
\sim C_0^{-2} \lambda_0^{-1} (H_0 t),
\end{equation}
where we used Eq. (\ref{eq:solution_phi}) and (\ref{eq:C0}).
For small $r$, the location of the brane $z_0$ is small, then $L$ is large. For
large $r$, $z_0$ is large, then $L$ is small.  

First let us examine the spectrum of the zero mode. 
Fig.2 shows the amplitude of the zero-mode $C_0$ for $\Delta=-2.645$. 
The amplitude is enhanced for large $r$. This results can be 
understood as follows \cite{trans, Frolov}. The effective 4D Planck 
scale $M_{p,eff}$ on the brane is given by
\begin{equation}
M_{p,eff}^2=M_5^3 L(t) = \left( \frac{M_p}{C_0} \right)^2 (H_0 t), 
\end{equation}
where $\kappa^2 =M_5^{-3}$ and $M_p$ is the Planck scale for static brane 
$M_p^2 = M_5^3 \lambda_0^{-1}$.
On the other hand, the Hubble scale at time $t$ is given by
$\dot{\alpha} =2/ (3\Delta+8)t$.
Then the vacuum expectation value of the 4D scalar field fluctuations 
$\varphi_4$ at the horizon might be
\begin{equation}
\frac{\langle \varphi_4^2  \rangle}{M_{p,eff}^2} \sim 
\frac{\dot{\alpha}^2}{M_{p,eff}^2}=
\frac{C_0^2}{M_p^2} 
\left(\frac{2}{3 \Delta+8} \right)^2 \frac{1}{H_0 t^3}.
\end{equation}
Now let us consider a mode with 3-wavenumber $p$. This mode crosses the 
horizon at $t=t_p$ where $p/e^{\alpha} \dot{\alpha} \vert_{t=t_p} =1$.
Then $t_p$ is given by \begin{equation}
H_0 t_p \sim  \left( \frac{p}{H} \right)^{-\frac{3 \Delta+8}{3(\Delta+2)}}.
\end{equation}
Thus amplitude of fluctuations with $p$ at the horizon is   
\begin{equation}
\left. \frac{\langle \varphi_4^2  \rangle}{M_{p,eff}^2}
\right \vert_{t=t_p} \sim 
C_0^2 \left( \frac{H^2}{M_p^2} \right) 
\left( \frac{p}{H} \right)^{\frac{3 \Delta+8}{\Delta+2}}.
\end{equation}
For a massless mode, this amplitude is frozen at super-horizon
scales. Indeed this is nothing but the dimension-less 
quantity defined by
\begin{equation}
\frac{ \langle   \varphi^2 \rangle_0}{M_5^3} \sim 
\frac{C_0^2 \lambda_0 H^2}{M_5^3}\left( \frac{p}{H} \right)^{\frac{3 \Delta+8}{\Delta+2}}
\sim \left. \frac{\langle \varphi_4^2  \rangle}{M_{p,eff}^2} 
\right \vert_{t=t_p} .
\end{equation}
For small $r$, $C_0$ becomes unity, then $M_{p,eff} \to M_{p}$.
Thus the standard 4D theory is recovered. On the other hand,
for large $r$, $C_0$ becomes large , then $M_{p,eff}$ becomes
small. Then the amplitude of the quantum fluctuations is enhanced.
Hence we conclude that the enhancement of the 0-mode amplitude for large 
$r$ is the result of the smallness of the effective 4D Planck scale at the inflation 
due to the small effective size of the extra-dimension .

\begin{figure}[t]
\epsfxsize=9cm
\centerline{\epsfbox{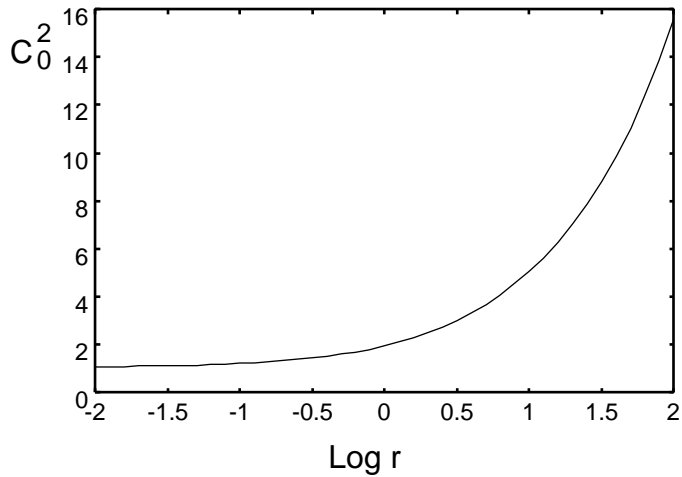}}
\caption{A factor $C_{0}^{2}$ in the zero mode amplitude (\ref{eq:massless}) as a
function of $r$ (\ref{eq:HL}). Here $\Delta = -2.645$.}
\label{fig:massless}
\end{figure}

\begin{figure}[t]
\epsfxsize=9cm
\centerline{\epsfbox{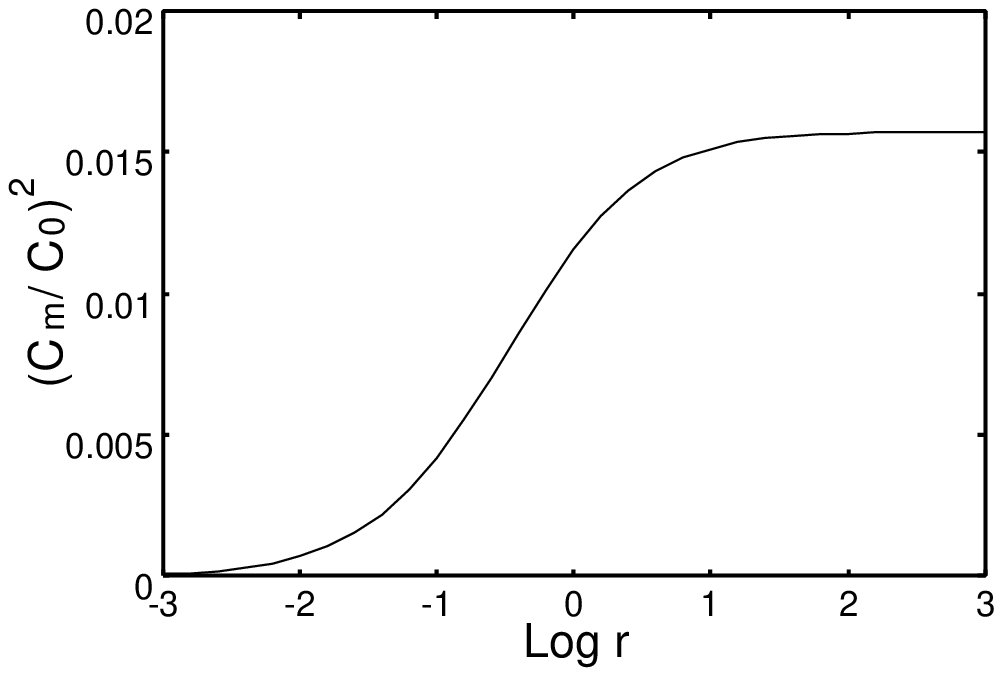}}
\caption{The ratio of the massive mode amplitude to the zero mode amplitude 
($C_{m}^{2} / C_{0}^{2}$).
Here $\Delta = -2.645$.}
\label{fig:ratio}
\end{figure}

Fig. \ref{fig:ratio} is the ratio $C_{m}^{2} / C_{0}^{2}$ as a function of $r$
for $\Delta=-2.645$ as an example. 
As is expected, the contribution of massive modes increases with $r$ but 
this ratio is at most$\sim 0.016$. This is due to the existence of the 
mass gap. For large $r$, $z_0$ becomes large. In the effective 
potential (\ref{eq:effective}), the term proportional to ${\cal H}(z)^{-2}$
which arises from the curvature of the bulk spacetime becomes irrelevant. 
The contribution of the massive modes are determined only 
by the mass gap which is independent of $r$. Indeed, for 
large $r$, $\vert \psi_m(z_0) \vert^2/\vert \psi_0(z_0) \vert^2$ approaches 
to Eq. (\ref{eq:ratio}) which is the result for no potential term
proportional to ${\cal H}(z)^{-2}$ (\ref{eq:nobarrier}).
The massive modes are too heavy to be excited by quantum fluctuations
thus the amplitude of the massive modes are suppressed. In addition,
the amplitudes of massive modes do not freeze out at super-horizon scales
and oscillate with decreasing amplitude. Then the spectrum 
becomes blue because the modes with small $p$ stay at super-horizon
scales longer times than the modes with large $p$. This decreasing 
of the amplitude does not depend on the mass for $m > -H/(\Delta+2)$, 
thus all massive modes have the same spectrum.

\section{Primordial fluctuations in bulk inflaton model \label{section:spectrum}}

Now we can compute the spectrum of primordial fluctuations generated
in the bulk inflaton model.
The amplitude of the CMB anisotopy caused by scalar perturbations 
is determined by the curvature perturbations ${\cal R}_c$ defined by 
\begin{equation}
{\cal R}_c \equiv \frac{\dot{\alpha}}{\dot{\phi}} \delta \phi_f.
\end{equation}
The amplitude of the scalar perturbation is then given by
\begin{equation}
\langle {\cal R}_c^2 \rangle = 
\left(\frac{\dot{\alpha}}{\dot{\phi}} \right)^2 \langle \delta \phi_f^2 \rangle
=\left(\frac{1}{3 \sqrt{2} b} \right)^2 \kappa^2  \langle \varphi^2 \rangle.
\end{equation}
On the other hand, the amplitude of the tensor perturbations is given by
\begin{equation}
\langle {h_{ij}^2} \rangle = 8 \kappa^2 \langle \varphi^2 \rangle,
\end{equation}
from Eq. (\ref{eq:tensor}). Here we take into account the polarization.
At large scales, the zero mode contribution is dominant. Thus we concentrate 
our attention on the zero mode contribution. Because the amplitude of the 
zero-mode is frozen at super-horizon scales, we expect that the amplitude
evaluated at horizon during inflation is directly related to the 
large scale CMB anisotropy.
The zero-mode contribution to the perturbations is written as 
\begin{eqnarray}
A_S^2 &=& \langle {\cal R}_c^2 \rangle_{0-mode} \sim  \left(\frac{1}{3 \sqrt{2} b} \right)^2 C_0^2
\left(\frac{H}{M_p} \right)^2 \left(\frac{p}{H} \right)^{\frac{3 \Delta+8}{\Delta+2}}, \nonumber\\ 
A_T^2 &=& \langle h_{ij}^2 \rangle_{0-mode} \sim 8 C_0^2  
\left(\frac{H}{M_p} \right)^2   
\left(\frac{p}{H} \right)^{\frac{3 \Delta+8}{\Delta+2}},
\end{eqnarray}
where Eq. (\ref{eq:massless}) and $\kappa^2=M_5^{-3}$, $M_p^2=M_5^3 \lambda_0^{-1}$ were
used.
The scalar field $\phi(t)$ can be regarded as a 4D scalar field
if we define $\phi_4=\lambda_0^{-1/2} \phi$ and $V_4(\phi_4) \propto M_p^2 \lambda_0^2 \delta
e^{-2 \sqrt{2} b \phi_4/M_p}$. Then using these effective 4D scalar field, the 
relative amplitude of the scalar and tensor perturbations is written as
\begin{equation}
\frac{A_T^2}{A_S^2} \sim M_p^2 \left(\frac{V_4'}{V_4} \right)^2.
\end{equation}
It would be interesting to compare our model with an inflation model driven 
by inflaton $\phi_b$ on the brane with potential $U_b(\phi_b)$ \cite{MWBH}. 
We assume that 
the bulk is AdS spacetime. Then it was shown that relative amplitude
of the scalar and tensor perturbations is given by
\begin{equation}
\frac{A_T^2}{A_S^2} \sim M_p^2 \left(\frac{U_b'}{U_b} \right)^2
\left(\frac{\lambda}{U_b} \right),
\end{equation}
for $U_b/\lambda \gg 1$ where $\lambda$ is the tension of the brane.
Thus for a given potential, the relative amplitude is suppressed for 
$U_b/\lambda \gg 1$.
This is because the enhancement of the scalar perturbations due to 
the rapid expansion $H \propto U_b$ is stronger than the enhancement 
of tensor perturbations due to the normalization in the bulk $C_0^2$. 
However in our model, the enhancement of the scalar perturbation 
is not caused by the unconventional expansion law but is caused by 
the normalization in the bulk $C_0^2$. Thus such a suppression is absent. 

Let us consider possible constraints on model parameters in our model. 
In our model there are three parameters; dilaton coupling $b$, 
tension of the brane $\lambda_0$ which determines the effective size 
of the extra-dimension at late times and $\delta$ which determines 
the energy scale of the inflation. Note that the scale of the
5D Planck scale $M_5$ is determined by $\lambda_0$ by 
$M_p^2 = M_5^3 \lambda_0^{-1}$ with $M_p =10^{19}$Gev. The dilaton 
coupling determines the spectrum index and relative amplitude
of the scalar and tensor perturbations. 
These quantities will be strongly tested in future observations
and these two observations give consistency check of our model. 
The COBE normalization of the amplitude gives one constraint on
the parameters $\lambda_0$ and $\delta$. Thus one free parameter 
is left, which can be taken as $r$. The parameter $r$ determines the 
cut-off frequency of the gravitational waves and the contribution
of the massive modes at high frequencies. It is quite challenging 
problems to observe stochastic gravitational waves at such high 
frequencies. Furthermore, the detailed spectrum of the gravitational 
waves may depend on the way to end the inflation and 
the subsequent evolution of the universe after inflation. 
Yet, it is certainly interesting issues to determine $r$ by the 
observations of gravitational waves.

\section{Discussions \label{section:discussion}}
In this paper we proposed an inflationary brane model driven by a bulk 
inflaton with exponential potential. A family of exact solutions that describe
power-law inflation on the brane was obtained. We calculated the primordial fluctuations in this model.
Tensor and scalar perturbations were shown to be described by
5D massless scalar field. Then the quantization of the 5D massless field
in this background was done. Using this result, the specturm of the 
primordial fluctuations were calculated. 

We should mention several open questions in our model. First, we need
some mechanism to end the inflation and stabilize the dilaton. 
Although we expect the large scale perturbations are insensitive to the 
details of these mechanism and our results will not be changed, 
they are important problems to be solved. Secondly, 
in order to connect our result with observations accurately, we should
know the subsequent evolution of the perturbations 
\cite{Mukohyama,KodamaIshibashiSeto,Maartens,Langlois2,Branden,KoyamaSoda2}. 
It has been shown that at large scales ${\cal R}_c$ remains constant
if we can neglect the dark radiation \cite{Large}. 
The tensor perturbation $h_{ij}$ also remains constant at large scales.
Thus the evolutions of ${\cal R}_c$ and $h_{ij}$
are very simple and our results can be directly used at the horizon re-enter.
However, in order to know the CMB anisotropy, we should know not only
the curvature perturbations ${\cal R}_c$ but also the anisotropic stress
induced by bulk gravitational fields which is measured by the difference
between $\Phi$ and $\Psi$. The evolution of this anisotropic stress is still 
uncertain. The evolution of the gravitational waves inside the horizon is also 
unknown. These issues deserve further investigations. 

\section{Acknowledgments}

We would like to thank Daisuke Ida, Tetsuya Shiromizu and Takashi Torii
for fruitful discussions. To complete this work, the discussions during 
the YITP workshop on Extra dimension and Braneworld was useful. 
KK's work and KT's work are supported by Grant-in-Aid 
for JSPS Fellows.

\end{document}